\pdfoutput=1
\documentclass[traditabstract]{aa} % for the abstract without structuration 
\usepackage{graphicx}
\usepackage{txfonts}
\begin{document}

   	\title{Galactic kinematics with modified Newtonian dynamics}
%	\subtitle{}
%	\titlerunning{}
%

   \author{O.   Bienaym\'e\inst{1}   \and   B.   Famaey\inst{2}   \and
X. Wu\inst{3} \and H.S. Zhao\inst{3} \and D. Aubert\inst{1} }

   \authorrunning{O. Bienaym\'e et al}
   \offprints{bienayme@astro.u-strasbg.fr}
%   \mail{}
%
   \institute{
Universit\'e de Strasbourg, CNRS, Observatoire Astronomique, France
\and
Institut d'Astronomie et d'Astrophysique, Universit\'e Libre de Bruxelles, Belgium
\and
SUPA, School of Physics and Astronomy, University of St. Andrews, UK}
   \date{27/03/2009 }
   \abstract{  We   look  for  observational   signatures  that  could
   discriminate  between  Newtonian   and  modified  Newtonian  (MOND)
   dynamics  in  the  Milky  Way,  in  view of  the  advent  of  large
   astrometric and spectroscopic  surveys. Indeed, a typical signature
   of MOND  is an  apparent disk of  ``phantom" dark matter,  which is
   uniquely correlated with the visible disk-density distribution. Due
   to this phantom  dark disk, Newtonian models with  a spherical halo
   have different  signatures from MOND  models close to  the Galactic
   plane.   The  models  can   thus  be  differentiated  by  measuring
   dynamically (within Newtonian dynamics) the disk surface density at
   the solar radius, the radial  mass gradient within the disk, or the
   velocity ellipsoid tilt angle  above the Galactic plane.  Using the
   most realistic possible  baryonic mass model for the  Milky Way, we
   predict that,  if MOND applies, the local  surface density measured
   by a Newtonist will be approximately 78 $M_\odot/{\rm pc}^2$ within
   1.1~kpc  of  the  Galactic  plane, the  dynamically  measured  disk
   scale-length will be  enhanced by a factor of  1.25 with respect to
   the visible disk  scale-length, and the local vertical  tilt of the
   velocity ellipsoid at 1~kpc above the plane will be approximately 6
   degrees.  None of these tests can be conclusive for the present-day
   accuracy of Milky Way data, but they will be of prime interest with
   the advent of large surveys such as GAIA.

{keywords: gravitation -- Stars: kinematics -- Galaxy: fundamental parameters --
Galaxy: kinematics and dynamics -- Galaxy: structure} }
   \maketitle
%
%________________________________________________________________
%
%%%%%%%%%%%%%%%	
%

\section{Introduction}

Our  understanding  of  Galactic  stellar populations  and  kinematics
achieves regular  progress with the introduction of  new large surveys
of  stars  with photometry,  distances,  radial  velocities or  proper
motions,  enabling us  to  probe the  6-dimensional  space of  stellar
positions and velocities (e.g., Perryman et al. \cite{hip}; Hog et al. ÷\cite{hog00};
Nordstr\"om et al. \cite{no04}; Famaey et al.  \cite{f05}; Zwitter et al.  \cite{zwi08}).
   
For instance,  the shape of  the Galactic 3-dimensional  potential has
been probed by,  e.g., the orbits of the  Sagittarius stream (Ibata et
al \cite{iba01}; Helmi \cite{helmi}; Read \& Moore \cite{RM05}; Haghi et
al. \cite{hag06}), the tidal tails  of Palomar 5 (Grillmair \& Dionatos
\cite{GD06}; Grillmair  \& Johnson \cite{GJ06}), or  the kinematics of
halo stars.   In the  solar neighbourhood, the  potential can  also be
analysed by measuring the Galactic escape speed of high velocity stars
(e.g.,   Smith  et   al.   \cite{smi07};   Famaey,  Bruneton   \&  Zhao
\cite{fbz07}),  the force  perpendicular to  the Galactic  plane (e.g.,
Oort \cite{oor60}; Cr\'ez\'e  et al \cite{cre98}; Siebert et al.   \cite{sie03}; Nipoti et al.
\cite{nip07}; Holmberg  \& Flynn \cite{hol00},  \cite{hol04}), the coupling between  the three
components  of the  velocity distribution  in the  solar neighbourhood
(e.g., Bienaym\'e \cite{bie99}), or  the orientation of the velocity ellipsoid
above  the   Galactic  plane  (Ollongren  \cite{oll62};  Hori   \&  Liu  \cite{hor63};
Lynden-Bell \cite{lyn62}; Siebert  et al. \cite{sie08}). In the  future, much progress
is  still expected to  be made  in our  understanding of  the Galactic
potential with the advent of the JASMINE and GAIA missions.

This mapping of the Galactic  potential is of great importance because
it holds the key to our  understanding of the behaviour of dark matter
on  galactic scales.   Nowadays, the  dominant paradigm  is  that dark
matter is  made of non-baryonic  weakly-interacting massive particles,
called ``cold  dark matter" (CDM).   However, on galactic  scales, the
observations  appear to be  at variance  with a  sizeable list  of CDM
predictions (see, e.g., Famaey  et al. \cite{fam07}).  The observed conspiracy
between the  mass profiles of baryonic  matter and dark  matter at all
radii  in spiral  galaxies (e.g.,  Famaey  et al.  \cite{fam07}) rather  lends
support  to  modified  Newtonian  dynamics  (MOND,  Milgrom  \cite{mil83}),  a
paradigm  postulating  that   for  accelerations  below  $a_0  \approx
10^{-10} {\rm  m} \, {\rm  s}^{-2}$ the true  gravitational attraction
approaches  $(g_N  a_0)^{1/2}$, where  $g_N$  is  the usual  Newtonian
gravitational field. Without resorting  to CDM, this paradigm is known
to reproduce galaxy scaling relations,  as well as the rotation curves
of individual galaxies  ranging over five orders of  magnitude in mass
(e.g., Sanders \& McGaugh \cite{San02}). In particular, the kinematic analysis
of  tidal dwarf galaxies  by Bournaud  et al.  (\cite{bou07}) is  difficult to
explain within the classical CDM  framework, while it is in accordance
with MOND  (Gentile et  al. \cite{Gen07}).  On  the other hand,  some problems
arise with  this paradigm in the subgalactic  and extragalactic scales
(e.g., Zhao \cite{Z05}; Angus et al.  \cite{ang07}; see e.g., Zhao \cite{Z08}; Angus \cite{ang09},
and Bruneton et al.  \cite{br09} for possible solutions).

The present  study examines possible observational  signatures of MOND
 gravity in the Milky Way,  which could be detected (or rejected) with
 the advent  of large Galactic  surveys. This is in  direct continuity
 with the works  of Famaey \& Binney (\cite{fam05}); Nipoti  et al. (\cite{nip07}); Wu
 et al. (\cite{wu08}),  and McGaugh (\cite{mcg08}). We build  these predictions using
 the MOND Milky Way model of Wu et al.~(\cite{wu08}), and refer the reader to
 this paper  for a full  description of the model\footnote{We  use the
 model  labelled  "MOND $g_{\rm  ext}=0.1a_0$"  in  Wu  et al.  (\cite{wu08}),
 meaning that  the modulus of the external  gravitational field acting
 on the Milky Way is chosen  to be $a_0/100$.}. This model is based on
 one of the most realistic  possible baryonic mass models of the Milky
 Way, the Besan\c{c}on model (Robin et al.~\cite{rob03}).

Once the gravitational potential of  the model is known, one can apply
the  Newtonian Poisson  equation to  recover the  density distribution
that would  have yielded this potential within  Newtonian dynamics. In
this  context,  MOND  predicts   a  disk  of  ``phantom"  dark  matter
(Sect.~2),  allowing us  to differentiate  the MOND  prediction from
these  of  a  Newtonian  model  with  a dark  halo  by  (i)  measuring
dynamically (within  Newtonian dynamics)  the disk surface  density at
the solar radius (Sect.~3), (ii) measuring the radial dynamical mass
gradient  within   the  disk  (Sect.~4),  or   (iii)  measuring  the
velocity-ellipsoid tilt angle above the Galactic plane (Sect.~5). We
show  that none  of these  tests are  yet conclusive  with present-day
accuracy,  but  they are  extremely  useful  to future  high-precision
astrometric and spectroscopic surveys.

\section{The ``phantom" disk of MOND}

Bekenstein \& Milgrom (\cite{bek84})  proposed a modification of the Newtonian
dynamics to reproduce the flat rotation curve of disk galaxies without
the  introduction  of enormous  amounts  of  dark  matter on  galactic
scales.   They   modified  the   Poisson  equation  by   relating  the
gravitational potential $\Phi$ to the mass density $\rho$ according to
\begin{equation}
\nabla\cdot  [ \,\mu (|| \nabla \Phi || / a_0) \, \nabla \Phi \,] = 4\,\pi\,G\,\rho .
\label{equ:bek}
\end{equation}
In  the  case of  high  acceleration,  the  function $\mu=1$  and  the
equation  becomes  the  usual  Poisson  equation.   The  case  of  low
acceleration is  explained and  described in a  long series  of papers
(see for instance Milgrom \cite{mil02}). At low accelerations (called the deep
MOND regime), when $||\nabla \Phi || \ll a_0$, the systems differ most
from Newtonian ones: assuming $\mu(t)\sim t$, the equation becomes:
\begin{equation}
 \nabla \cdot  [ \, || \nabla   \Phi || \,  \nabla \Phi \,] = 4\,\pi\,G\,a_0\,\rho.
\label{equ:deep}
\end{equation}
Spherical density-potential pairs are  easily built and some flattened
spheroids can be numerically described (see for instance Ciotti et al.
\cite{cio06}). In  the Appendix, we  develop a new numerical  potential solver
for this equation.

Within  MOND,   one  finds  that   any  disk  or   spheroidal  density
distribution of  finite mass produces  a spherical potential  at large
radius (see Appendix). It is the Mondian explanation of the dark mater
halo  detected by  a  Newtonist. However,  a  disky distribution  also
produces a supplementary disk  potential (Milgrom \cite{mil83}, \cite{mil01}; see also
Appendix).

Once the MOND  gravitational potential is known, one  can always apply
the  Newtonian  Poisson  equation   to  it,  to  recover  the  density
distribution $\rho_{newton}$  corresponding to the  given potential in
Newtonian dynamics.  Thus, how  can we differentiate\footnote{ MOND is
compatible  (i.e., not  falsified)  with recent  determination of  the
Galactic halo flattening, for  instance deduced from the kinematics of
the  Sagittarius stream  (Ibata  et  al \cite{iba01};  Read  \& Moore  \cite{RM05}).}
Newtonian  from  Mondian dynamics,  since  the  difference of  density
$\rho_{newton}-\rho_{mond}$ (given  by the two theories  from the same
potential) can  always be attributed by  a Newtonist to  a dark matter
component (the  so-called ``phantom" dark matter of  MOND)? The answer
lies within the properties of the supplementary disk of ``dark matter"
that appears in the phantom density $\rho_{newton}-\rho_{mond}$.

The properties  of this phantom disk  are very specific  and unique to
MOND. Of course,  two different kinds of dark  matter components (with
distinct kinematics)  could explain both  the round and the  thin dark
components  (see  e.g.,  Read  et  al.   \cite{Read08}).   However,  if  future
observations  reveal  such  a  thin  dark-matter disk,  of  the  exact
charateristics predicted by MOND, they will clearly make MOND dynamics
more relevant to explaining Galactic dynamics, while failing to detect
it  could falsify  the  Bekenstein \&  Milgrom  (\cite{bek84}) formulation  of
MOND. Our goal hereafter is  to quantify the predicted effects of this
dark  matter  disk on  Galactic  kinematics  in  the Milky  Way.  This
specific  dark (phantom)  disk coud  be easily  identified  using GAIA
data; however, the exact predictions  of MOND depend on its ability to
describe  the visible  matter distribution.  For that  purpose, models
such  as the Besan\c  con model  are necessary  because it  takes into
account  important details  such as  the change  in both  radial scale
length, and the mass distribution  between thin, thick disks and other
baryonic components.

We  note that,  hereafter, when  we speak  about the  ``dynamical mass
distribution  in MOND", we  always speak  about the  mass distribution
that  would have  yielded the  MOND gravitational  potential  from the
Newtonian Poisson  equation, i.e., the baryonic mass  plus the phantom
dark mass.

\section{The local dynamical surface density in MOND}

Milgrom (\cite{mil83})  already established that  the phantom dark  disk would
    enhance  the  $K_{\rm  z}$  force perpendicular  to  the  Galactic
    plane. To be compatible with the analysis of Hipparcos data in the
    solar neighbourhood,  this enhancement must not be  too strong and
    must  not imply  an  extremely  massive unseen  disk.  Based on  a
    Galactic model, Nipoti et al.  (\cite{nip07}) determined that at the solar
    position  MOND almost doubles  the $K_{\rm  z}$ force  produced in
    Newtonian gravity by  the baryonic disk. For this  model (based on
    model   1  of   Dehnen  and   Binney,  \cite{deh98}),   the  baryonic-disk
    surface-mass  density is  $\Sigma_0$=43 M$_\odot$  pc$^{-2}$. When
    the disk is embedded in  a spherical dark-matter halo, the surface
    density becomes 65 M$_\odot$ pc$^{-2}$ (see their Figure 5), while
    for MOND (baryons+phantom) it becomes 85 M$_\odot$ pc$^{-2}$. This
    corresponds  respectively to  an  increase of  51  percent and  98
    percent compared to the surface density of the baryonic disk.

Using the  Milky Way MOND model  of Wu et  al.~(\cite{wu08}), we found
instead  an increase  of 57,  62, and  66 percent  at $R=7.5$,  8, and
8.5~kpc  respectively (78~$M_\odot/{\rm  pc}^2$ at  $R=8.5$~kpc). This
difference with  Nipoti el al.  (\cite{nip07})  results from different
baryonic-mass concentrations and  different local $\mu$ values between
the models.  This dynamical surface density of 78~$M_\odot/{\rm pc}^2$
must be compared with the  $K_{\rm z}$ force determination at 1.1 kpc,
$74\pm6$M$_\odot$  pc$^{-2}$  by  Holmberg  \&  Flynn  (\cite{hol04}).
Considering   the  various   uncertainties,  these   measurements  are
compatible both  with the Newton+spherical halo  and MOND Besan\c{c}on
models. Indeed,  these accurate determinations of the  $K_{\rm z}$ are
based  on clearly defined,  but small,  samples of  a few  hundreds of
stars.  Due to the small size of these samples, only one parameter can
be  efficiently  recovered:  the  surface density  below  some  height
$\bar{z}$, the mean  distance of the samples from  the Galactic plane.
This  does not  break  the  degeneracy between  two  effects: a  small
flatenning  of the  halo and  a small  increase of  the  baryonic disc
density.   These  two effects  can  compensate  without modifying  the
surface density measured at  $|z|=$1.1\,kpc.  In contrast, the $K_{\rm
z}(z)$ shape at smaller $z$ (and also the Oort limit) will change.  In
the future,  large samples  from RAVE and  very large ones  from GAIA,
will make it  possible to recover precisely $K_{\rm  z}$ versus z, and
to   determine  dynamically   the  detailed   vertical   mass  density
distribution $\rho_{\rm dyn}(z)$.

\section{The dynamical disk scale-length in MOND}

A second and different test concerns the prediction of the $K_{\rm z}$
force perpendicular  to the  Galactic plane at  various Galactocentric
radii,    to   measure   dynamically    the   scale-length    of   the
(visible+phantom) disk  mass distribution and  to compare this  to the
visible disk  scale-length. The interesting  property of this  test is
that  only the  gradients in  the  density distributions  of both  the
visible  and  visible+phantom  disks   have  to  be  measured.   These
gradients are  certainly easier  to measure than  the true  total mass
distribution.

In Fig.~1,  we plot (dashed  line) the radial density  distribution of
the  baryonic  matter  in  the  Galactic plane  for  the  Besan\c  con
model.   By  adopting   the  MOND   gravitational  potential   (Wu  et
al.  \cite{wu08}), we  then determine  the  baryonic+phantom dynamical
density (solid line)  that would be inferred by  a Newtonist measuring
the $K_{\rm z}$  force perpendicular to the Galactic  plane at various
Galactocentric radii. A factor  of $\sim$1.25 is predicted between the
two  scale  lengths, determined  between  5 kpc  and  10  kpc from  the
Galactic centre.

Since the Galactic thin disc dominates the disk mass distribution with
a radial  scale length of 2.5\,kpc  (Robin et al.  \cite{rob03}), the expected
disk  dynamical  scale-length  differs significantly,  3.1\,kpc,  
between measurements by a Newtonist and by a MONDian. This measurement
will be a test of prime  interest when data from large surveys such as
GAIA or JASMINE become available.

\begin{figure}[!htbp]
\resizebox{\hsize}{!}{      
\includegraphics[width=9.cm,height=7.cm]{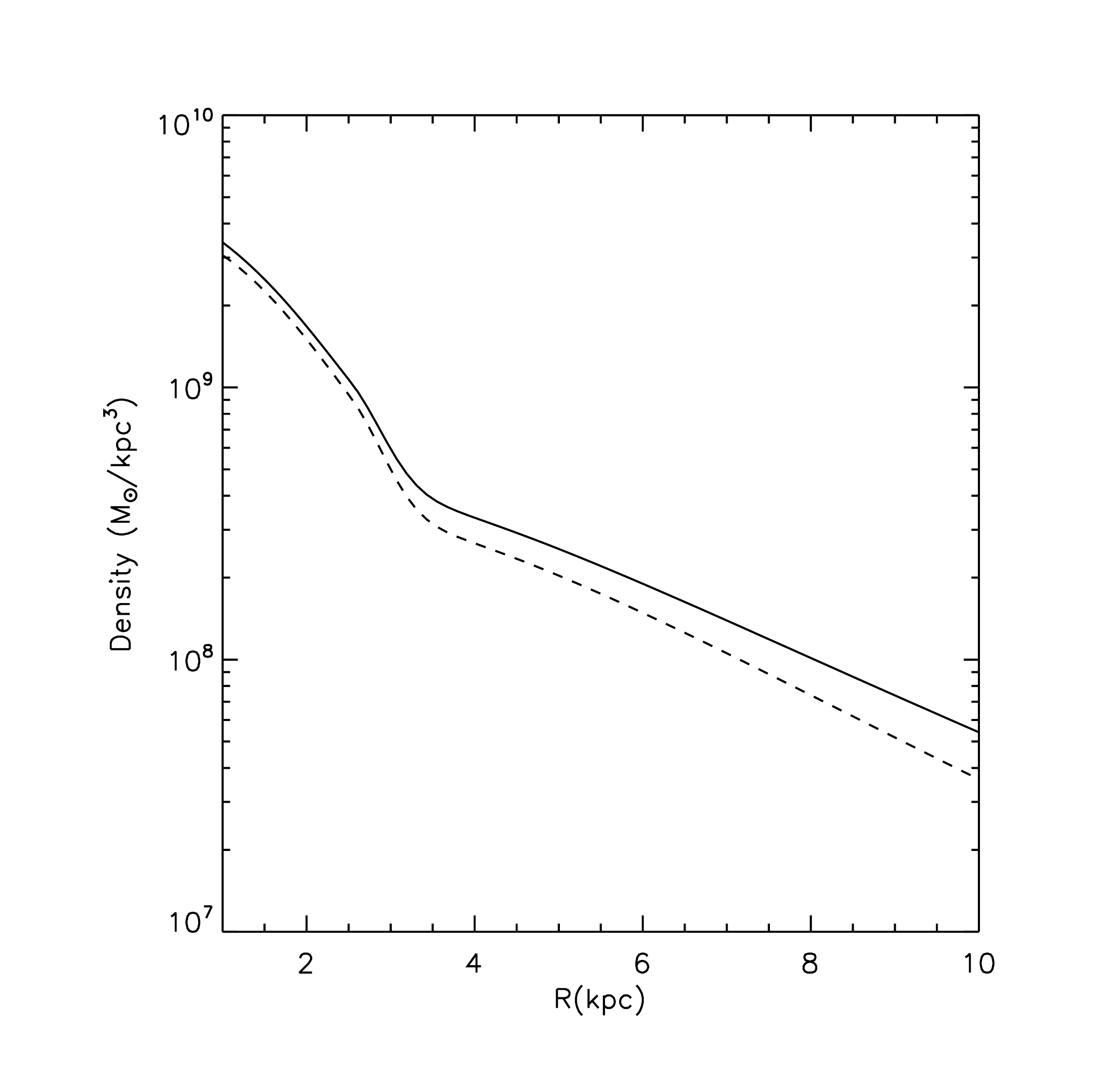}
}

\caption{{\bf   Dashed   line}:   The  radial   density   distribution
$\rho(R,z=0)$  of the baryonic  matter within  the Besan\c  con model.
{\bf Solid  line}: The baryonic+phantom density in  the Galactic plane
from the Besan\c con MOND model.
}
\label{f:besancon}
\end{figure}

\section{The tilt angle of the velocity ellipsoid in MOND}

A third new and interesting test  of MOND relies on the measure of the
vertical  tilt of the  stellar velocity  ellipsoid above  the Galactic
plane (e.g., Siebert et al.  \cite{sie08}). At any position above the Galactic
plane,  solar  neighbourhood stars  are  distributed along  isodensity
ellipsoids in  the $(UVW)$-velocity-space  (where $U$ is  the velocity
towards  the Galactic  centre, $V$  the velocity  in the  direction of
Galactic rotation, and $W$ the  vertical veloicty, all with respect to
the  Sun). At  a  given position  $z$  above the  Galactic plane,  the
inclination angle $\delta$ of the ellipsoid minor axis with respect to
the $W$ axis (see Fig.~6 of  Siebert et al. \cite{sie08}) is intimately linked
to  the gravitational  potential.  In the  $UW$-plane,  this angle  is
defined by
\begin{equation}
{\rm tan} 2\delta = \frac{2\sigma^2_{UW}}{\sigma^2_U - \sigma^2_W},
\end{equation}
where   $\sigma^2_{UW}$,  $\sigma^2_U$,   and  $\sigma^2_W$   are  the
second-order velocity distribution moments. Cuddeford \& Amendt (\cite{cu91})
analysed,  to  the  fourth  order,  the  consecutive  moments  of  the
Boltzmann equation  by developing these moments by  expanding a Taylor
series. Combining these moment equations, they derived expressions for
the velocity  moments and obtained  an approximate expression  for the
tilt  angle  $\delta$,   which  depend  only  on  the   shape  of  the
gravitational potential  (but is  exact only for  completely separable
axisymmetric  potentials). More  precisely, the  ellipsoid inclination
close to  the Galactic plane is  related to the  disk scale-length and
also  to  the dark-matter  halo  flattening  (Bienaym\'e \cite{bie09}).   This
dependence of the vertical tilt of the velocity ellipsoid on the shape
of the potential has also been checked by numerical simulations (e.g.,
Siebert et al. \cite{sie08}).

\begin{figure}[!htbp]
\resizebox{\hsize}{!}{
\includegraphics[angle=270]{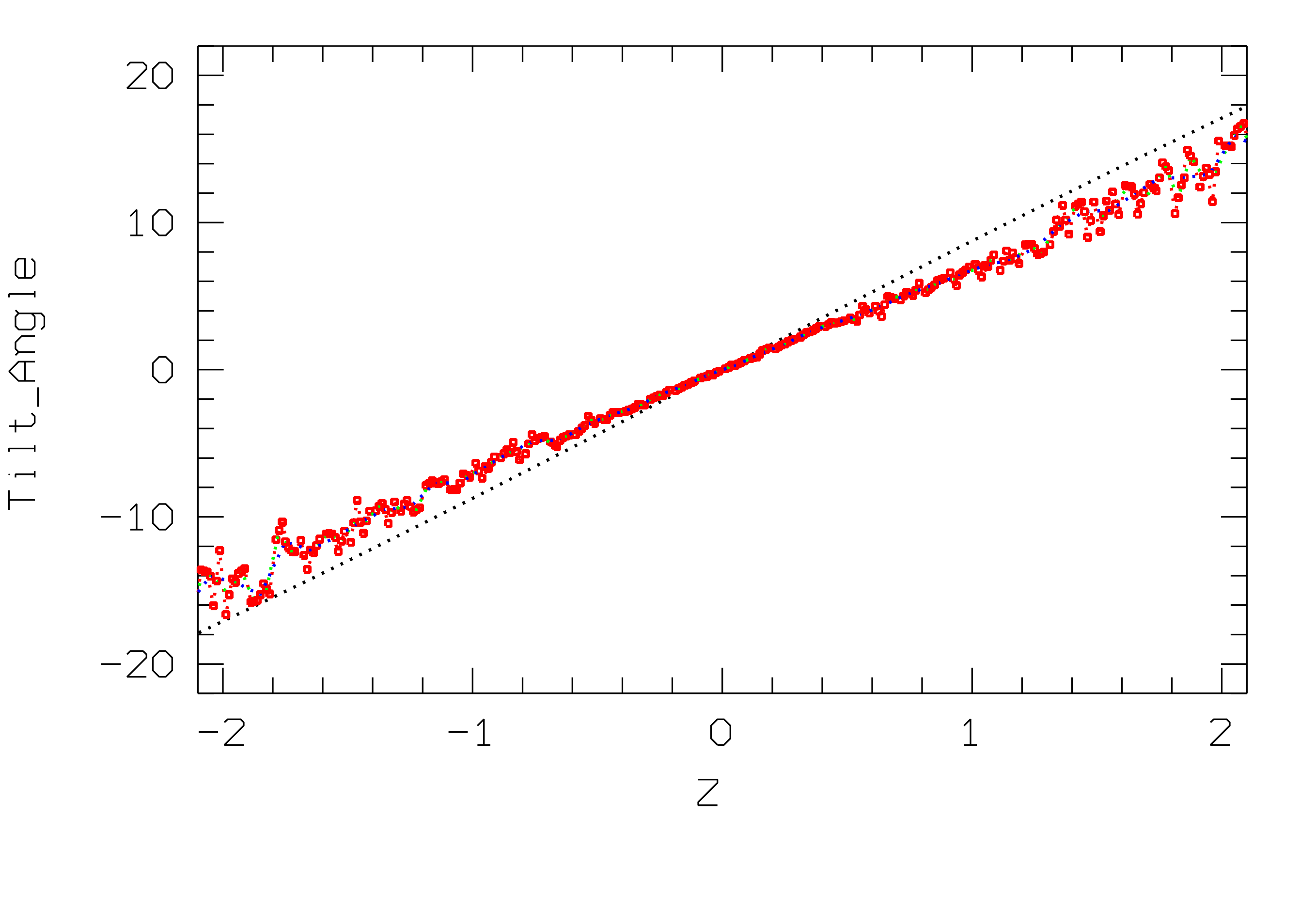}
\includegraphics[angle=270]{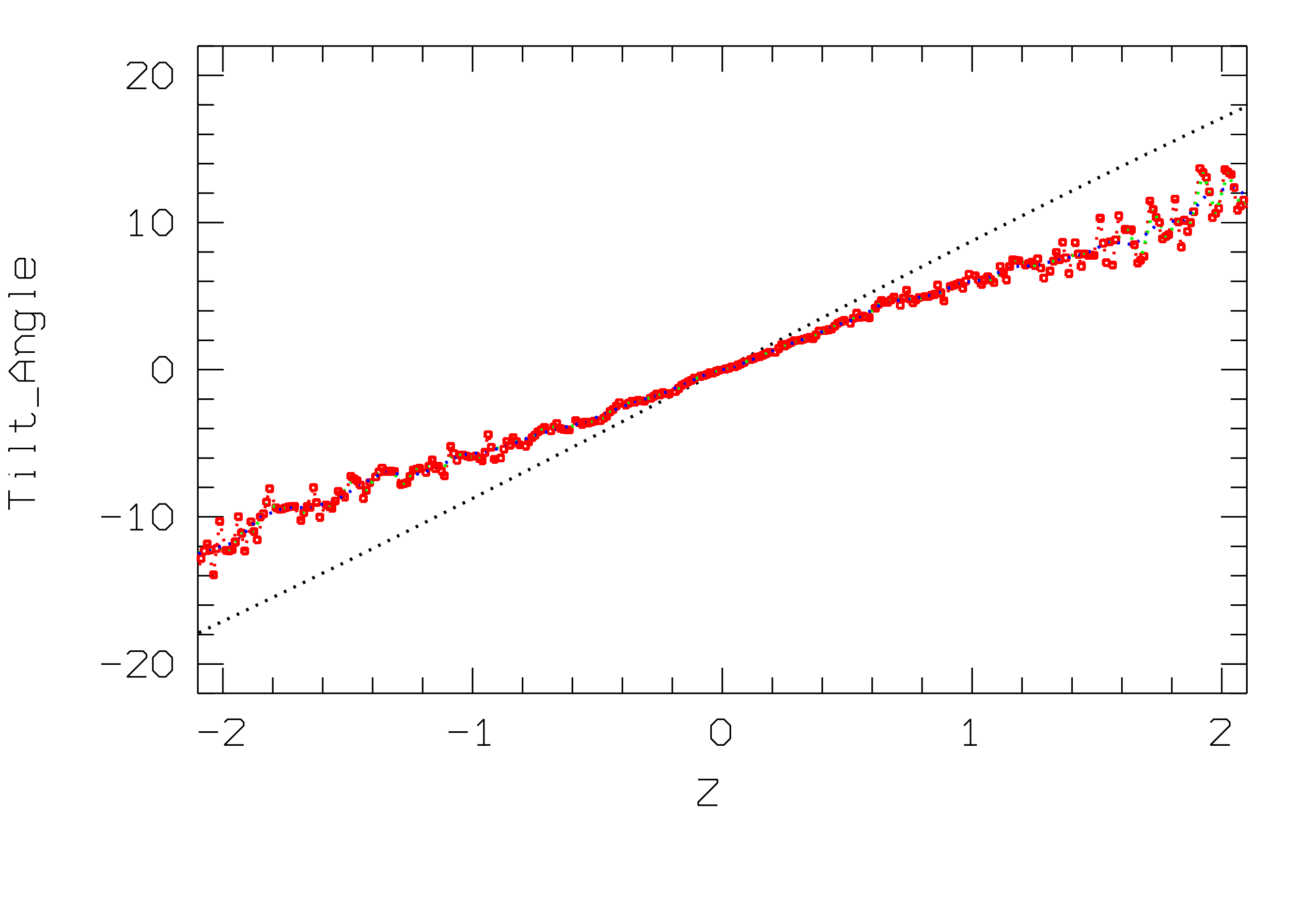}
}
\resizebox{\hsize}{!}{
\includegraphics[angle=270]{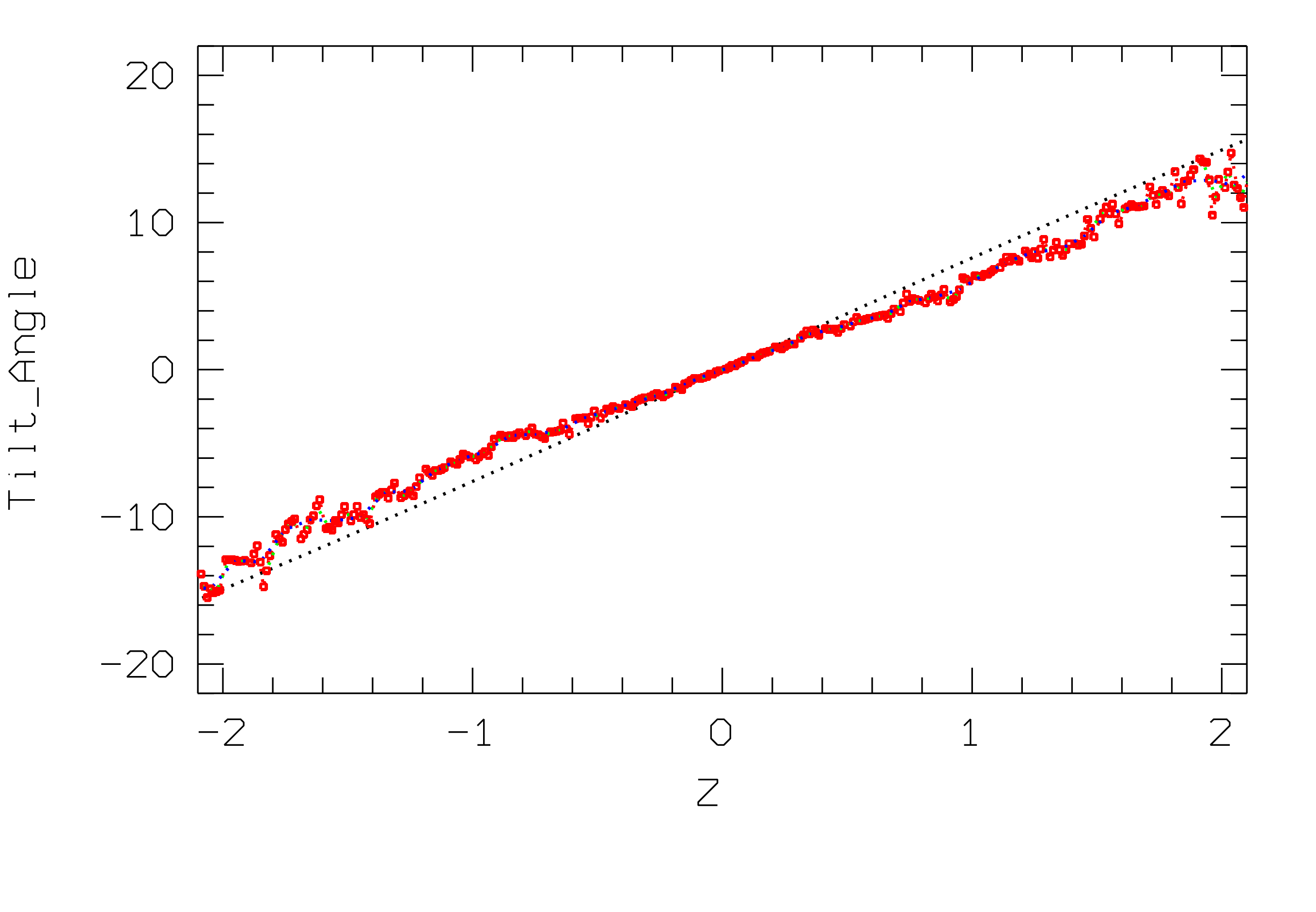}
\includegraphics[angle=270]{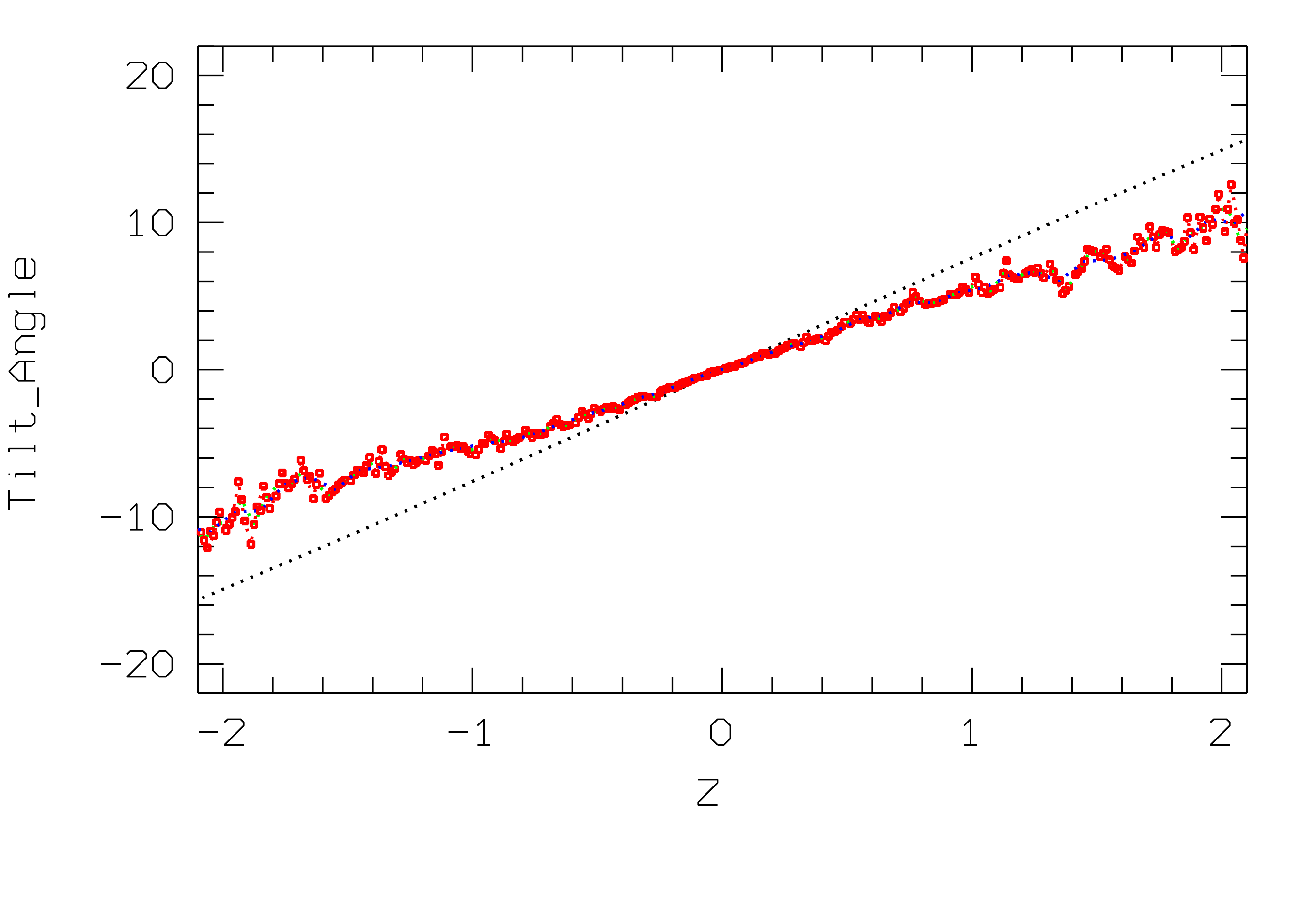}
}
\resizebox{\hsize}{!}{
\includegraphics[angle=270]{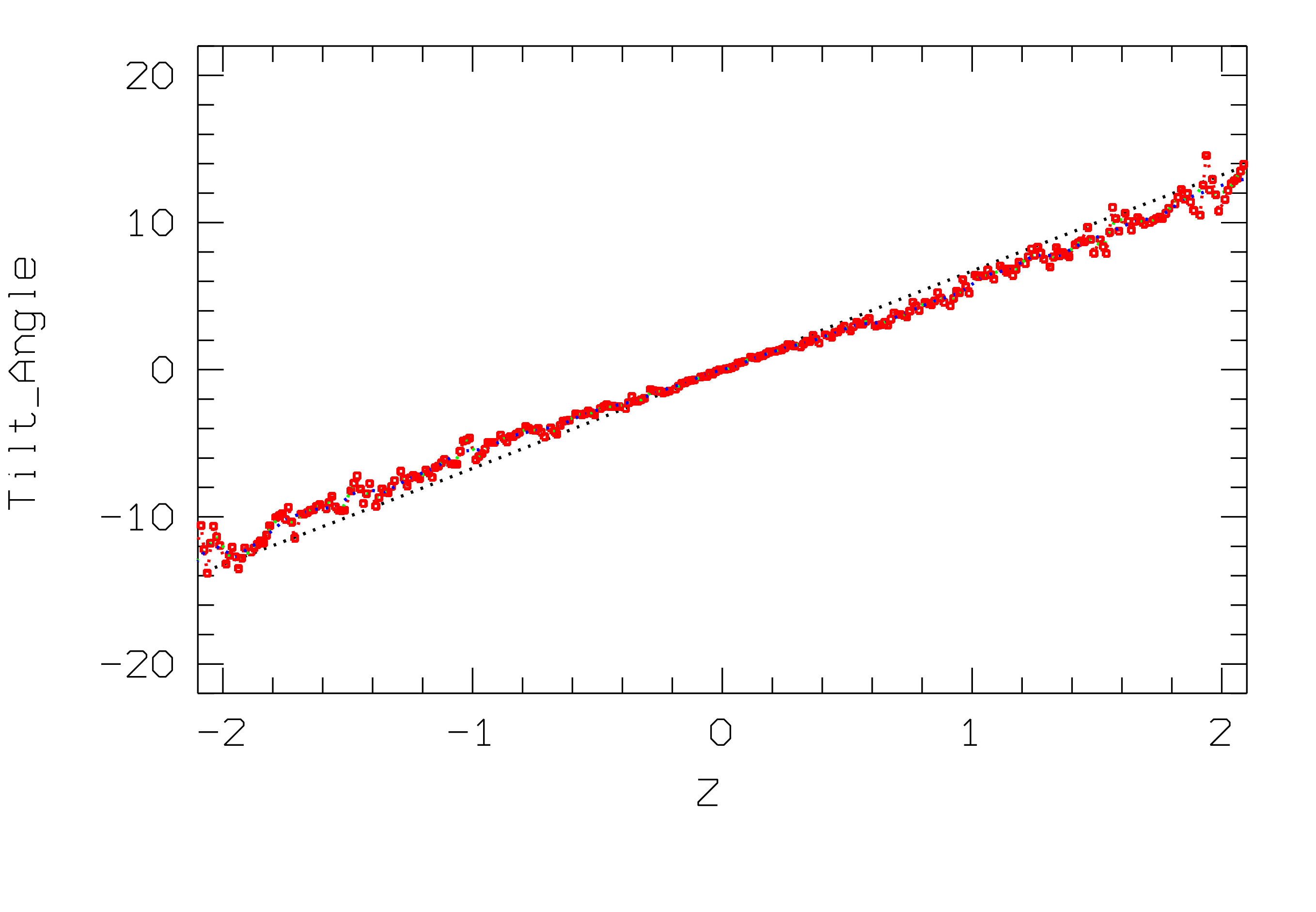}
\includegraphics[angle=270]{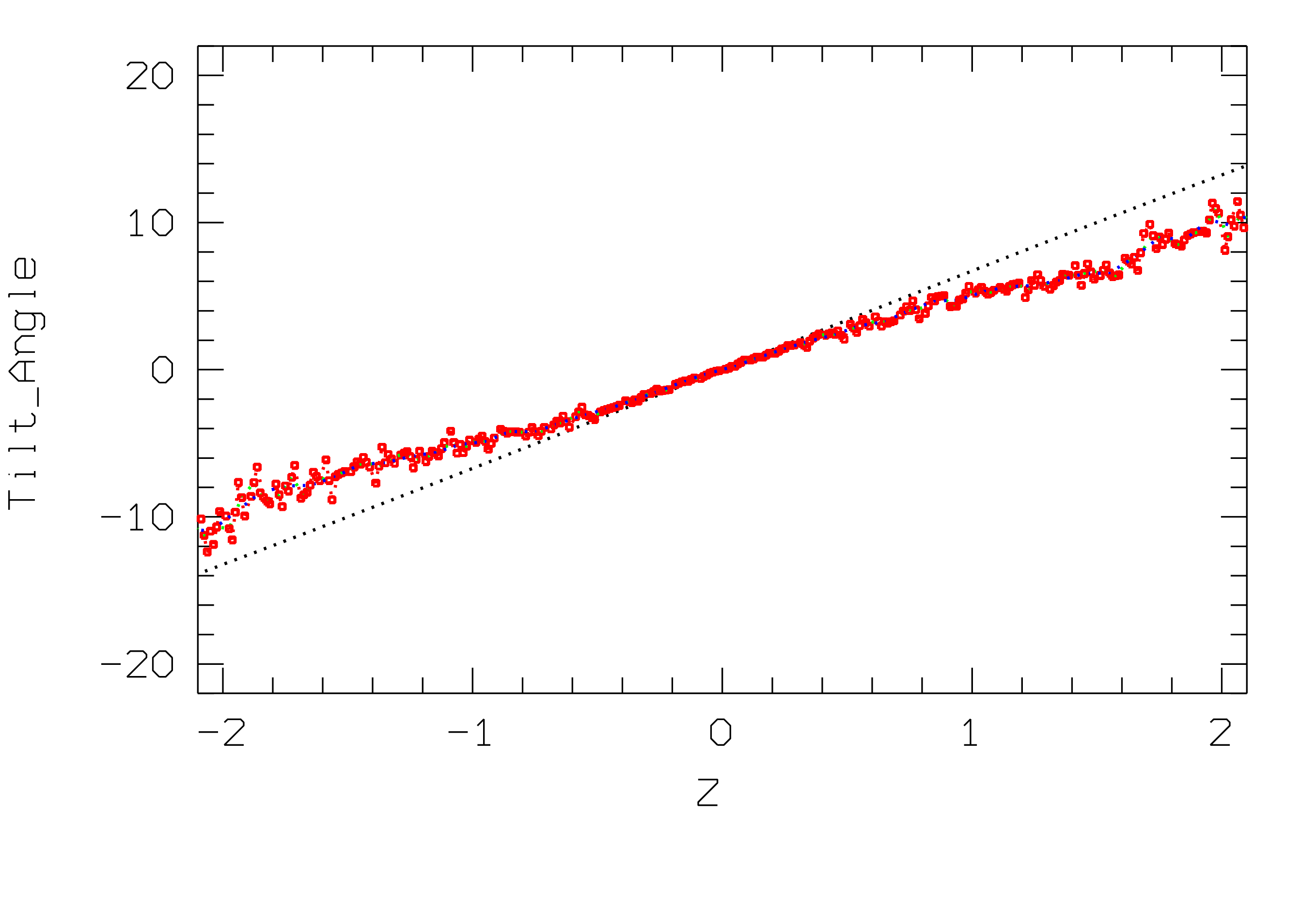}
}
\caption{Velocity ellipsoid  tilts in degrees versus $z$  at $R$= 6.5,
  7.5  and  8.5  (top  to  bottom)  (the  dotted  line  would  be  the
  inclination within  a spherical potential).  {\bf  Left:} within the
  Newtonian potential of  the Besan\c con Galactic model  (Robin et al
  \cite{rob03}),  {\bf Right:}  within  the MONDian  potential of  the
  Besan\c con model (Wu et al \cite{wu08}). }
\label{f:ExpoTilt}
\end{figure}

Here, we  determine the vertical tilt  angle as a function  of $z$ (at
the Galactocentric  radius of the  Sun) for the Newtonian  Besan\c con
model embedded in a dark halo (Robin et al. \cite{rob03}), and compare
this  inclination  with  that  obtained  for the  MOND  model  (Wu  et
al.  \cite{wu08}).  For this,  we  follow  Siebert  et al.~(\cite{sie08}),  and
compute  the  inclination of  the  velocity  ellipsoid by  integrating
orbits in the meridional plane  (we note however that the Besan\c{c}on
model  is non-axisymmetric,  so that  we average  it).  We  sample the
initial  conditions  from  a  Shu  distribution  function  (Bienaym\'e
\cite{bie99}),  yielding an  orbit library  of over  20  million orbits.
Each  orbit was  integrated  over  120 rotations,  using  a 4th  order
Runge-Kutta algorithm.  We then randomly selected  1 phase-space point
per  orbit in  the  last 80  rotations.  From there,  we computed  the
inclination  angle  of the  velocity  ellipsoid  at different  $(R,z)$
positions.   Figure~\ref{f:ExpoTilt} displays  the  resulting vertical
tilt  angle (calculated  with  Eq. 3) as  a  function of  $z$ at  three
different  Galactocentric  locations  for  both the  Newtonian  (+dark
matter) and  MONDian Besan\c{c}on models.  This tilt  is also compared
with the  tilt expectation from  a purely spherical  potential (dotted
lines on  Fig.~\ref{f:ExpoTilt}): clearly, the flattening  of the MOND
potential  due  to the  presence  of a  phantom  dark  disk makes  the
inclination  angle lower than  for a  purely spherical  potential, and
lower  than  for  the  Newtonian  Besan\c{c}on  model  embedded  in  a
spherical halo. The tilt differences are most significant in the inner
part of  the Galaxy. At  the solar position, however,  the differences
remain small when $|z|\le $ 1kpc, and become significant only at $z=2$
kpc (a difference  of, 3.5 and 2 degrees,  respectively at $R$=7.5 and
8.5 kpc,  the inclination  being smaller within  the MOND  model).  At
$z=1$~kpc  and $R=8.5$~kpc  (the  solar position  in  the Besan\c  con
model), where the strongest observational constraints on the tilt have
been obtained, both models predict  an inclination of 6 degrees, which
agrees with  the observational determination of  $7.3\pm 1.8$ degrees,
obtained   with   a   RAVE   sample   of   580   stars   (Siebert   et
al.  \cite{sie08}). Other observational  results on  the measure  of the
vertical  tilt remain a  bit controversial  and conflicting  (Fuchs et
al. \cite{fuchs09}; Smith et al.   \cite{smi09}), and will require more (and
more  precise)  data to  avoid  any  potential  bias, and  confirm  or
disprove the sphericity of the potential.

\begin{center}
\begin{table*}
   \caption{Values predicted from the  Besan\c{c}on MOND model as seen
   by  a Newtonist  compared to  observations, for  the  local surface
   density and the tilt of the velocity ellipsoid.}

\begin{tabular}{ |l |c| c| }
\hline  
   & Besan\c{c}on  MOND  & Observations  \\
\hline  
$\Sigma_\odot(z=1.1  \,  {\rm  kpc})$  &  78  $M_\odot/{\rm pc}^2$  &  74  $\pm$  6
$M_\odot/{\rm pc}^2$ (Holmberg \& Flynn \cite{hol04}) \\
\hline
Tilt at $z=1$~kpc & 6 degrees & 7.3 $\pm$ 1.8 degrees (Siebert et al. 2008) \\
\hline

\end{tabular}
\end{table*}
\end{center}

A measurement of  the necessary accuracy will possibly  be achieved at
the solar Galactic radius with  large RAVE samples in the near future,
and  with GAIA data  in a  few more  years.  It  will be  important to
account   for   secondary   effects   (see,  e.g.,   Famaey   et   al.
\cite{fam05}), such as streams, resonances, and the bar influence, and
to correctly  take into account their  impact on the  vertical tilt of
the  ellipsoid.   However,  we  note,  that, in  the  immediate  solar
neighbourhood, only  the Hyades are  clearly identified in  the $(UW)$
distribution that looks extremely well phase mixed, in contrast to the
$(UV)$  distribution  where  many  substructures (from  resonances  or
non-stationnarity) have  been identified.  This  is also true  for the
$(U,W)$ distributions  of distant samples at 1\,kpc  from the Galactic
plane   (Soubiran   et   al.    \cite{sou08}  and   Siebert   et   al.
\cite{sie08}).  The effect of the bar and specific vertical resonances
needs  to  be understood  in  greater  detail  since they  can  modify
significantly  the  vertical-tilt orientation  (Oll\'  e \&  Pfenniger
\cite{oll98}).  However, this family  of orbits should also be visible
as substructures in the $(UV)$ plane.

%%%% figures des inclinaisons

\section{Conclusions}

In addition to  the predictions made by Wu  et al.~(\cite{wu08}) about
the rotation curve and the escape speed, MOND also makes very specific
predictions  about Galactic kinematics  related to  the presence  of a
disk of  phantom dark matter.   These predictions will be  tested with
future large surveys:

(i)  By measuring the  $K_{\rm z}$  force: at  the solar  radius, MOND
predicts a 60 percent enhancement  in the dynamical surface density at
1.1~kpc relative to the baryonic surface density, a value not excluded
by current data.  This enhancement would become more apparent at large
Galactic   radii,   where   the  stellar-disk   mass-density   becomes
negligible.

(ii)  By determining  dynamically  the scale-length  of the  disk-mass
 density-distribution.  This  scale length  is a factor  of $\sim$1.25
 larger  than the  scale-length of  the visible  stellar disk  if MOND
 applies.   This  test  could  be  applied  with  existing  RAVE  data
 (Steinmetz et al. \cite{ste06}; Zwitter et al.  \cite{zwi08}), but the accuracy
 of available proper motions still limits the possibility of exploring
 the gravitational forces too far from the solar neighbourhood.

(iii)  By  measuring the  velocity  ellipsoid  tilt  angle within  the
 meridional  Galactic  plane.  This  tilt  is  different  for the  two
 dynamics in the inner part of the Galactic disk. However, the tilt of
 about  6 degrees  at  z=1~kpc at  the  solar radius  agrees with  the
 determination  of  $7.3\pm  1.8$   degrees  obtained  by  Siebert  et
 al. (\cite{sie08}). The difference between  MOND and a Newtonian model with a
 spherical halo becomes significant at z=2~kpc.

These tests  of gravity could be  applied with future  GAIA or JASMINE
data that will allow us to reconstruct the 3-dimensional gravitational
potential of  the Galaxy.  To assess the  values of the  {\it current}
local constraints, the  predictions of the Besan\c con  MOND model are
compared with the relevant observations in Table~1.

\begin{acknowledgements}

The  authors  gratefully acknowledge  Annie  Robin  for  her help.  BF
acknowledges financial support of the FNRS.

\end{acknowledgements}

%_______________________________________________________
%
% BIBLIOGRAPHIE
%_______________________________________________________
%\bibliographystyle{aa}

\begin{appendix}
\section{Full multi-grid MOND potential solver}

We developed a  new full multigrid (FMG) solver in  the same spirit as
  Brada    \&   Milgrom   (\cite{bra95})    and   Tiret    \&   Combes
  (\cite{tir07}). It solves Eq. (\ref{equ:bek}) by means of relaxation
  performed  on the  density field  sampled at  different resolutions.
  Large-scale feature  convergence is obtained from  a coarse sampling
  of the density, while high-frequency features result from fine grids
  (see e.g.,  Press et al.  \cite{NR02}).  Relaxation  is acheived
  with a 4 colours Gauss-Seidel, with 5 pre/post iterations.The solver
  is  materially accelerated  by  means of  Graphical Processor  Units
  (GPUs) using the CUDA API  developed by NVidia for its hardware. The
  Poisson equation  is solved in $\sim  1$ second over  a $256^3$ grid
  using a GeForce 8800 GTX device. Typical relative residuals obtained
  for the MOND Poisson equation are $0.4 \%$.

 We tested our potential solver with disks  of  finite
thickness, namely an exponential disk
(scale length  $l$=1, thickness $h=0.2$), and a  thickened Kuzmin disk
(scale length $a$=1, and thickness $h$=0.2):
\begin{eqnarray}
\rho_{exp}(R,z) = \exp\,(-R/l) \exp\,(-(z/h)^2)\\
\rho_{kuz}(R,z) =   \left( 1+(R/a)^2+(z/h)^2\right)  ^{-5/2} .
\label{equ:dens}
\end{eqnarray}

\begin{figure}[!htbp]
\resizebox{\hsize}{!}{
\includegraphics{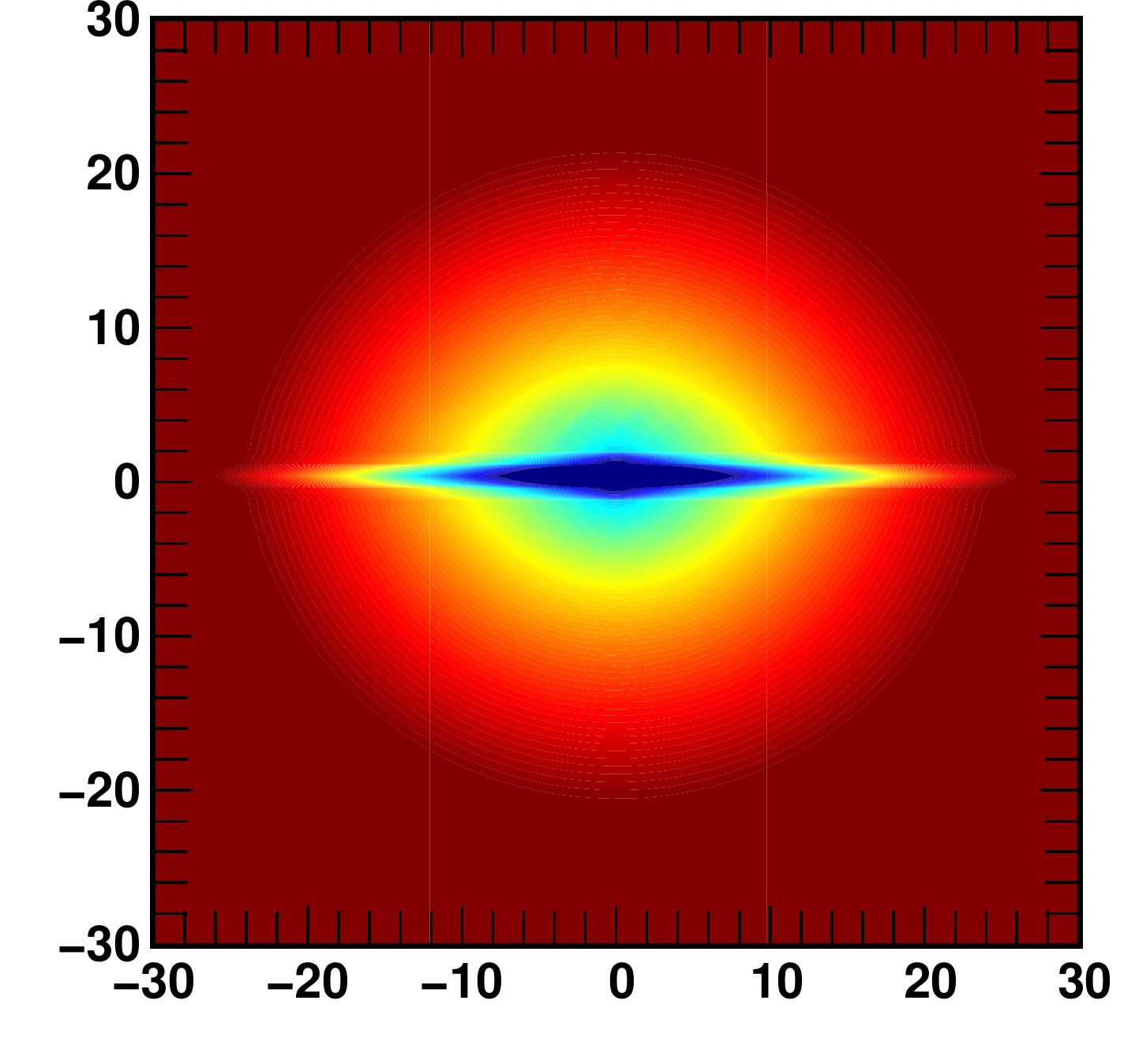}
}
\caption{ Effective volume mass-density measured by  a Newtonist from
the  deep-MOND  potential of  an exponential  disk  (scale length=2.5,
scale height=0.5).}
\label{f:mond-exp1a}
\end{figure}

\begin{figure}[!htbp]
\resizebox{\hsize}{!}{
\includegraphics{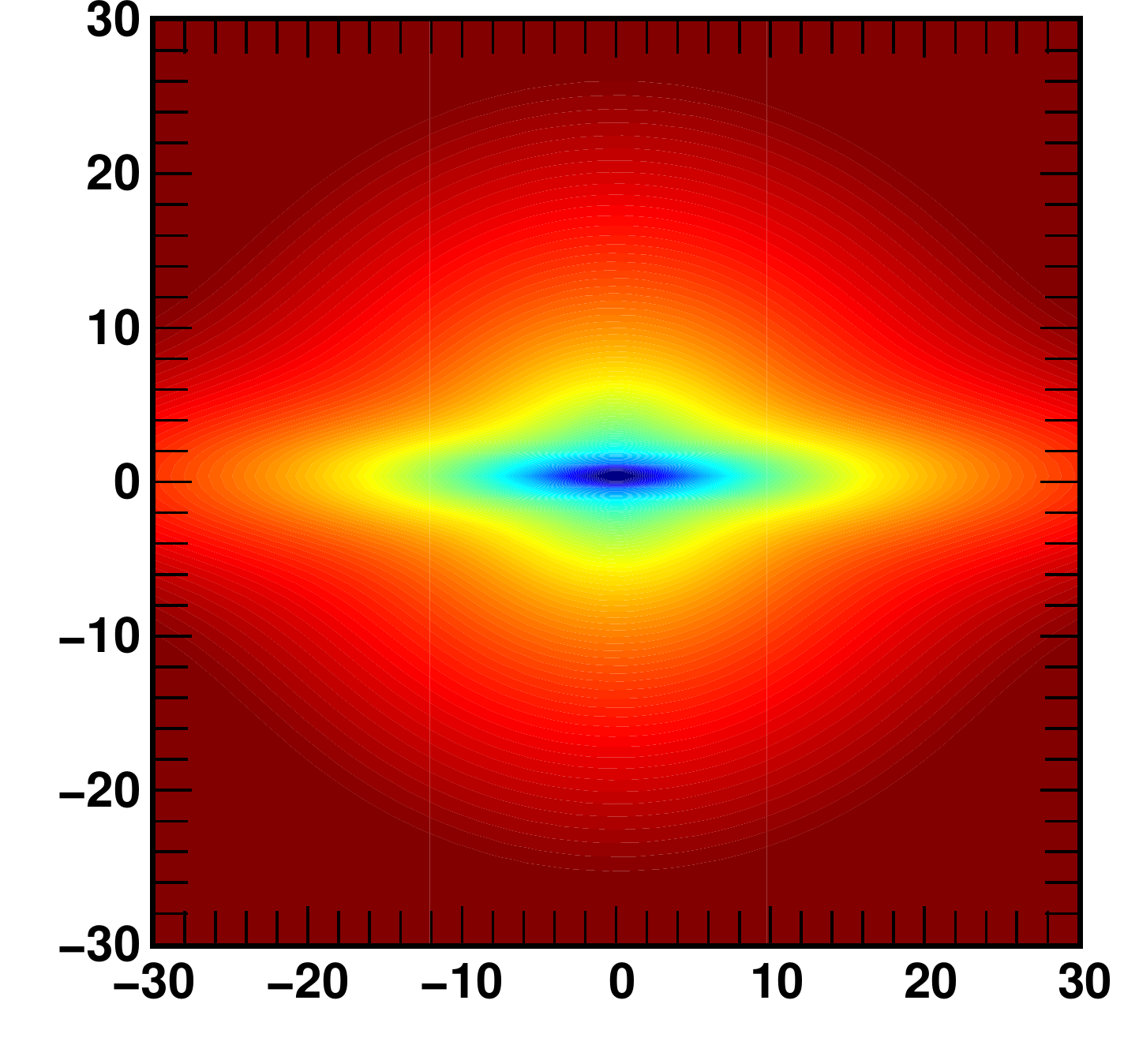}
}
\caption{Effective volume mass-density as measured by a Newtonist from
the deep-MOND  potential of a thickened Kumin  disk (scale length=2.5,
scale height=0.5).}
\label{f:mond-kuz1a}
\end{figure}

The  FMG  results  for  the effective  (visible+phantom)  density  are
  plotted in Figs.~\ref{f:mond-exp1a}  and \ref{f:mond-kuz1a}. We note
  that  the phantom  disk  clearly  appears in  the  outskirts of  the
  Galaxy.   The  FMG  results   were  previously  checked  by  solving
  Eq.\,(\ref{equ:deep})  using  the   FreeFem++  software,  a  partial
  differential  equation solver  (Hecht et  al.  \cite{hecht}).  For spherical
  potentials the accuracy of  the computed potential with FreeFem++ is
  excellent.   With  very flattened  systems  (axis  ratio 1/10),  the
  achieved  accuracy remains good  thanks to  an adaptive  grid within
  FreeFem++. For  instance, in the  case of the exponential  disk, our
  mean relative accuracy  is 1.5 per cent at radii  $R$ of between 0.4
  and 6 scale lengths, and lower than 5 per cent for radii between 0.2
  and 10 scale lengths.  Excellent  agreement is found between the FMG
  and  FreeFem++,  with relative  differences  of  smaller than  $1\%$
  within 10 scale lengths.

%%%%% Figure

\end{appendix}

\end{document}